%tbh.tex
%version 05.02.2001
%&Plain
  \input lanlmac.tex
\overfullrule=0pt
\input epsf.tex
%\draft
%\newcount\figno
\figno=0
\def\fig#1#2#3{
\par\begingroup\parindent=0pt\leftskip=1cm\rightskip=1cm\parindent=0pt
\baselineskip=11pt
\global\advance\figno by 1
\midinsert
\epsfxsize=#3
\centerline{\epsfbox{#2}}
\vskip 12pt
{\bf Fig. \the\figno:} #1\par
\endinsert\endgroup\par
}
\def\figlabel#1{\xdef#1{\the\figno}}
\def\encadremath#1{\vbox{\hrule\hbox{\vrule\kern8pt\vbox{\kern8pt
\hbox{$\displaystyle #1$}\kern8pt}
\kern8pt\vrule}\hrule}}

%macros
%
\font\zfont = cmss10 %scaled \magstep1
\font\litfont = cmr6

\def\bigone{\hbox{1\kern -.23em {\rm l}}}
\def\ZZ{\hbox{\zfont Z\kern-.4emZ}}
\def\hf{{\litfont {1 \over 2}}}

\def\p{\partial}

\def\X{\Xi}

\def\oo{\hat \omega   }

\def\o{\omega }

%%%%%%%%%%%%%%%%%%%ELLIPTIC%%%%%%%%%%%%%%%%%%%%%%%%%%%%%

%%%%%%%%%%%%%%%%%%%%%%%%%%%%%%%%%%%%%%%%%%%%%%%%%%%%%%%%%%%%%

%%%%%%%%%%%%%%%%%%%%% Calligraphic letters %%%%%%%%%%%%%%%%%%%%

   \def\CF {{\cal F}}

   \def\CO {{\cal O}}

   \def\CZ {{\cal Z}}
  % %%%%%%%%%%%%RRRRRRZZZZZZZZZZZZ%%%%%%%%%%%%%%%%%%%%%%%%%%%%

  %%%%%%%%%%%%RRRRRRZZZZZZZZZZZZ%%%%%%%%%%%%%%%%%%%%%%%%%%%%
   \def\R{\relax{\rm I\kern-.18em R}}
   \font\cmss=cmss10 \font\cmsss=cmss10 at 7pt
   \def\Z{\relax\ifmmode\mathchoice
   {\hbox{\cmss Z\kern-.4em Z}}{\hbox{\cmss Z\kern-.4em Z}}
   {\lower.9pt\hbox{\cmsss Z\kern-.4em Z}}
   {\lower1.2pt\hbox{\cmsss Z\kern-.4em Z}}\else{\cmss Z\kern-.4em
   Z}\fi}
   %%%%%%%%%%%%derivatives%%%ETC%%%%%%%%%%%%%%%%%%%%%%%%%%%
   \def\p{\partial}
   
   \def\11{1\!\! 1}

    \def\hepth{ {\tt hep-th/}}

%%%%%%%%%%%%%%%% VARIABLES %%%%%%%%%%%%%%%%%%%%%%%%%%%%
%%%%%%%%%%%%%%%% VARIABLES %%%%%%%%%%%%%%%%%%%%%%%%%%%%

\def\tl{{\tilde \lambda}}

%%%%%%%%%%%%%%%%%%% + and - %%%%%%%%%%%%%%%%%%%%%%%%%

%%%%%%%%%%%%%%%%%%% coordinates %%%%%%%%%%%%%%%%%%%%%%
\def\xp{z_{_{+}}}
\def\xm{z_{_{-}}}
\def\xpm{z_{_{\pm}}}

\def\zp{z_{_{+}}}
\def\zm{z_{_{-}}}
\def\zpm{z_{_{\pm}}}

\def\bop{\omega^{}_{_{+}}}
\def\bom{\omega^{}_{_{-}}}
\def\bopm{\omega^{}_{_{\pm}}}

\def\oo{\o_{(0)}}

%%%%%%%%%%%%%%%%%%% constants %%%%%%%%%%%%%%%%%%%%%%%%%%
\def\tp{t_1}
\def\tm{t_{-1}}
\def\tpm{t_{\pm 1}}
\def\tmp{t_{\mp 1}}

%%%%%%%%%%%%%%%%%%% matrices %%%%%%%%%%%%%%%%%%%%%%%%%%

\def\Mpm{M_{\pm }}

%%%%%%%%%%%%%%%%%%% wave functions %%%%%%%%%%%%%%%%%%%%%%%%

\def\dpi{{1\over 2\pi}}
\def\oR{{1\over R}}
\def\X{X}
\def\ea{e^{-{2\over R}\X}}
\def\eb{e^{-{1\over R^2}\X}}
\def\ec{e^{-2{1-R\over R^2}\X}}
\def\ef{e^{-{1\over R}\X}}
\def\eg{e^{-{1-R\over R^2}\X}}

\def\fs{\rm ^{_{Fermi\ sea}}}

%%%%%%%%%%%%%%%%%REFERENCES%%%%%%%%%%%%%%%%%%%%%%%
%%%%%%%%%%%%%%%%%REFERENCES%%%%%%%%%%%%%%%%%%%%%%%

\def\np#1#2#3{{\it Nucl. Phys.} {\bf B#1} (#2) #3}
\def\pl#1#2#3{{\it Phys. Lett. }{\bf B#1} (#2) #3}
\def\prl#1#2#3{{\it Phys. Rev. Lett.}{\bf #1} (#2) #3}
\def\physrev#1#2#3{{\it Phys. Rev.} {\bf D#1} (#2) #3}
\def\prb#1#2#3{{\it Phys. Rev.} {\bf B#1} (#2) #3}

\def\rmatp#1#2#3{{\it Rev. Math. Phys. }{\bf #1} (#2) #3}
\def\cmp#1#2#3{{\it Comm. Math. Phys.} {\bf #1} (#2) #3}
\def\mpl#1#2#3{{\it Mod. Phys. Lett. }{\bf #1} (#2) #3}
\def\ijmp#1#2#3{{\it Int. J. Mod. Phys.} {\bf #1} (#2) #3}
\def\lmp#1#2#3{{\it Lett. Math. Phys.} {\bf #1} (#2) #3}
\def\tmatp#1#2#3{{\it Theor. Math. Phys.} {\bf #1} (#2) #3}
\def\jhep#1#2#3{{\it JHEP} {\bf #1} (#2) #3}
\def\anp#1#2#3{{\it Annalen Phys.} {\bf #1} (#2) #3}
\def\hepth#1{{\tt hep-th/}#1}
\def\grqc#1{{\tt gr-qc/}#1}

%%%%%%%%%%%%% our %%%%%%%%%%%%%%%%%%%%%%%%%
\lref\HKK{ J. Hoppe, V. Kazakov, and I. Kostov,
``Dimensionally reduced SYM$_4$ as solvable matrix quantum
mechanics'', \np{571}{2000}{479},  \hepth{9907058}.}
\lref\NKK{V. Kazakov, I. Kostov and  N.  Nekrasov,
``D-particles, matrix integrals and KP hierachy'',
\np{557}{1999}{413},  \hepth{9810035}.}
\lref\KKK{V. Kazakov, I. Kostov, and D. Kutasov,
``A Matrix Model for the Two Dimensional Black Hole",
\np{622}{2002}{141}, \hepth{0101011}.}
\lref\AK{S. Alexandrov and V. Kazakov,
``Correlators in 2D string theory with vortex condensation'',
\np{610}{2001}{77}, \hepth{0104094}.}
\lref\IK{I. Kostov, ``String Equation for String Theory on a Circle'',
\np{624}{2002}{146}, \hepth{0107247}.}
\lref\AKK{S.Yu. Alexandrov, V.A. Kazakov, and I.K. Kostov,
``Time-dependent backgrounds of 2D string theory'',
\np{640}{2002}{119}, \hepth{0205079}.}
\lref\SAprog{S.Yu. Alexandrov, in progress.}

%%%%%%%%%%%%%%% matrix models %%%%%%%%%%%%%%%%%%%%%%
\lref\KAZMIG{V. Kazakov and A. A. Migdal, \np{311}{1988}{171}.}
\lref\BRKA{E. Brezin, V. Kazakov and Al. Zamolodchikov,
\np{338}{1990}{673}.}
\lref\PARISI{ G. Parisi, \pl{238}{1990}{209, 213}.}
\lref\GRMI{ D. Gross and N. Miljkovic, \pl{238}{1990}{217}.}
\lref\GIZI{P. Ginsparg and J. Zinn-Justin, \pl{240}{1990}{333}.}
\lref\BIPZ{ E. Brezin, C. Itzykson, G. Parisi, and J.-B. Zuber,
\cmp{59}{1978}{35}.}

%%%%%%%%%%%%%%%% vortex %%%%%%%%%%%%%%%%%%%%%%%%%%%%%
\lref\GRKL{D. Gross and I. Klebanov, \np{344}{1990}{475};
  \np{354}{1990}{459}.}
\lref\BULKA{D. Boulatov and V. Kazakov,
``One-Dimensional String Theory with Vortices as Upside-Down
Matrix Oscillator'', \ijmp{8}{1993}{809}, \hepth{0012228}. }

%%%%%%%%%%%%% Black hole %%%%%%%%%%%%%%%%%%%%%%%
\lref\DVV{R. Dijkgraaf, E. Verlinde, and H. Verlinde,
``String propagation in a black hole geometry'',
\np{371}{1992}{269}.}
\lref\FZZ{ V. Fateev, A. Zamolodchikov, and Al. Zamolodchikov,
{\it unpublished}.}
\lref\AKKII{ S. Yu. Alexandrov, V. A. Kazakov, I. K. Kostov,
{\it work in progress}. }
\lref\MUKHIVAFA{S. Mukhi and C. Vafa,
``Two dimensional black-hole as a topological coset model of
c=1 string theory'', \np{407}{1993}{667}, \hepth{9301083}.}
\lref\GOSHALVAFA{D. Ghoshal and C. Vafa,
``c=1 String as the Topological Theory of the Conifold'',
\np{453}{1995}{121}, \hepth{9506122}.}
\lref\KZ{V. A. Kazakov and A. Tseytlin,
``On free energy of 2-d black hole in bosonic string theory'',
\jhep{0106}{2001{}021}, \hepth{0104138}.}
\lref\TECHNER{J. Teschner,
``The deformed two-dimensional black hole'',
\pl{458}{1999}{257}, \hepth{9902189}. }
\lref\JAP{T. Fukuda and K. Hosomichi,
``Three-point Functions in Sine-Liouville Theory'',
\jhep{0109}{2001}{003}, \hepth{0105217}. }
\lref\GP{G. W. Gibbons and M. J. Perry,
``The Physics of 2-D stringy space-times'',
\ijmp{1}{1992}{335}, \hepth{9204090}. }
\lref\NP{C. R. Nappi and A. Pasquinucci,
``Thermodynamics of two-dimensional black holes'',
\mpl{A7}{1992}{3337}, \hepth{9208002}. }
\lref\KumVas{W. Kummer and D. V. Vassilevich,
``Hawking radiation from dilaton gravity in (1+1) dimensions:
A pedagogical review'',
\anp{8}{1999}{801}, \grqc{9907041}. }
\lref\DasMathur{S. R. Das and S. D. Mathur,
`'The Quantum Physics of Black Holes: Results from String Theory'',
{\it Ann. Rev. Nucl. Part. Sci.} {\bf 50} (2000) 153,
\grqc{0105063}. }
\lref\Unruh{S. R. Unruh,
`'Notes on black hole evaporation'',
\physrev{14}{1976}{870}.}

%%%%%%%%%%% Moore  %%%%%%%%%%%%%%%%%%%%%%
\lref\MP{G. Moore and M. Plesser,
``Classical scattering in 1+1 Dimensional string theory'',
\physrev{46}{1992}{1730}, \hepth{9203060}.}
\lref\MPR{G. Moore, M. Plesser, and S. Ramgoolam,
``Exact S-matrix for 2D string theory'',
\np{377}{1992}{143}, \hepth{9111035}.}
\lref\MOORE{G. Moore, ``Gravitational phase transitions
  and the sine-Gordon model", \hepth{9203061}.}
\lref\DMP{R. Dijkgraaf, G. Moore, and M.R. Plesser,
``The partition function of 2d string theory'',
\np{394}{1993}{356}, \hepth{9208031}.}

%%%%%%%%%%%%%%%%%%% scattering %%%%%%%%%%%%%%%%%%%%%%
\lref\POLCHINSKI{J. Polchinski, ``What is string theory'',
{\it Lectures presented at the 1994 Les Houches Summer School
``Fluctuating Geometries in Statistical Mechanics and Field Theory''},
\hepth{9411028}.}
\lref\KLEBANOV{I. Klebanov, {\it Lectures delivered at the ICTP
Spring School on String Theory and Quantum Gravity},
Trieste, April 1991, \hepth{9108019}.}
\lref\JEVICKI{A. Jevicki, ``Developments in 2D string theory'',
\hepth/9309115.}
\lref\HSU{E. Hsu and D. Kutasov, ``The Gravitational Sine-Gordon Model'',
\np{396}{1993}{693}, \hepth{9212023}.}
\lref\MSS{G. Moore, N. Seiberg, and M. Staudacher,
``From loops to states in 2D quantum gravity'',
\np{362}{1991}{665}. }

%%%%%%%%%%%%%%%%%%%%% Toda %%%%%%%%%%%%%%%%%%%%%%%%%%
\lref\JM{M. Jimbo and T. Miwa, ``Solitons and Infinite Dimensional
Lie Algebras'', {\it Publ. RIMS, Kyoto Univ.} {\bf 19}, No. 3
(1983) 943.}
\lref\Hir{R. Hirota, Direct Method in Soliton Theory {\it Solitons},
Ed. by R.K. Bullogh and R.J. Caudrey, Springer, 1980.}
\lref\UT{K. Ueno and K. Takasaki, ``Toda Lattice Hierarchy":
in `Group representations and systems of differential equations',
{\it Adv. Stud. Pure Math.} {\bf 4} (1984) 1.}
\lref\Takasak{K. Takasaki,
{\it Adv. Stud. Pure Math.} {\bf 4} (1984) 139.}
\lref\MukhiImbimbo{ C. Imbimbo and S. Mukhi,
``The topological matrix model of c=1 String",
\np{449}{1995}{553}, \hepth{9505127}.}
\lref\NTT{ T.Nakatsu, K.Takasaki, and S.Tsujimaru,
``Quantum and classical aspects of deformed $c=1$ strings'',
\np{443}{1995}{155}, \hepth{9501038}.}
\lref\Takebe{T. Takebe, ``Toda lattice hierarchy and conservation laws'',
\cmp{129}{1990}{129}.}
\lref\EK{T. Eguchi and H. Kanno,
``Toda lattice hierarchy and the topological description of the
$c=1$ string theory'', \pl{331}{1994}{330}, hep-th/9404056.}
\lref\Krichever{I. Krichever, {\it Func. Anal. i ego pril.},
{\bf 22:3} (1988) 37  (English translation:
{\it Funct. Anal. Appl.} {\bf 22} (1989) 200);
``The  $\tau$-function of the
universal Witham hierarchy, matrix models and topological field
theories'', {\it Comm. Pure Appl. Math.} {\bf 47} (1992),
\hepth{9205110}.}
\lref\TakTakb{K.~Takasaki and T.~Takebe,
``Quasi-classical limit of Toda hierarchy and W-infinity symmetries'',
\lmp{28}{93}{165}, \hepth{9301070}.}
\lref\orlshu{A. Orlov and E. Shulman, \lmp{12}{1986}{171}.}
\lref\Nakatsu{T. Nakatsu, ``On the string equation at $c=1$'',
\mpl{A9}{1994}{3313}, \hepth{9407096}.}
\lref\TakSE{K. Takasaki,
``Toda lattice hierarchy and generalized string equations'',
\cmp{181}{1996}{131}, \hepth{9506089}.}
\lref\TakTak{K. Takasaki and T. Takebe,
``Integrable Hierarchies and Dispersionless Limit'',
\rmatp{7}{1995}{743}, \hepth{9405096}.}

%%%%%%%%%%%%%%%%%% kaplya %%%%%%%%%%%%%%%%%%%%%%%%%
\lref\kkvwz{ I. Kostov, I. Krichever, M. Mineev-Veinstein,
P. Wiegmann, and  A. Zabrodin, ``$\tau$-function
for analytic curves", \hepth{0005259}.}
\lref\Zabrodin{ A. Zabrodin, ``Dispersionless limit of Hirota
equations in some problems of complex analysis'',
\tmatp{129}{2001}{1511}; \tmatp{129}{2001}{239}, {\tt math.CV/0104169}.}
\lref\bmrwz{ A. Boyarsky, A. Marshakov,  O. Ruchhayskiy,
P. Wiegmann, and  A. Zabrodin, ``On associativity equations in
dispersionless integrable hierarchies",
\pl{515}{2001}{483}, \hepth{0105260}.}
\lref\wz{P. Wiegmann and  A. Zabrodin, ``Conformal maps and
dispersionless integrable hierarchies", \cmp{213}{2000}{523},
\hepth{9909147}.}
\lref\FERTIGone{H.A. Fertig, \prb{36}{1987}{7969}. }
\lref\FERTIGtwo{H.A. Fertig, \prb{387}{1988}{996}. }
\lref\WIEGAGAM{O. Agam, E. Bettelheim, P. Wiegmann, and A. Zabrodin,
``Viscous fingering and a shape of an electronic droplet in
the Quantum Hall regime'', {\tt cond-mat/0111333}.}
\lref\mwz{ M.~Mineev-Weinstein, P.~B.~Wiegmann, A.~Zabrodin,
``Integrable Structure of Interface Dynamics'', \prl{84}{2000}{5106},
{\tt nlin.SI/0001007}.}

%%%%%%%%%%%%%%%%%%%%%%%%%%%%%%%%%%%%%%
\lref\WITTENGR{E. Witten, ``Ground Ring of two dimensional string theory'',
\np{373}{1992}{187}, \hepth{9108004}.}
\lref\MuchiImbimbo{C. Imbimbo
and S. Mukhi, ``The topological matrix model of c=1 String",
\hepth{9505127}.}

%%%%%%%%%%%%%%%%%%%%%%%%%%%%%%
   %%%%%%%%%%%%%%%%%%%%%%%%%%%%%%%%%%%%%%%%%%%%%%%%%%%%%%%%%%%%
%\rightline{SPHT-t02/158, LPTENS-02/***}
\Title{
%\rightline{hep-th/0210251}
}
%\vbox{\baselineskip12pt\hbox
%{SPHT-t00/123}\hbox{LPTENS-00/32}\hbox{EFI-2000-29} }}
{\vbox{\centerline{Thermodynamics of 2D string theory   }
%\centerline{   }
}}
%   \footnote{}{*optional footnote on title}
%
%
\centerline{Sergei Yu. Alexandrov,$^{123}$\footnote{$^{\ast}$}
{alexand@spht.saclay.cea.fr}
and Vladimir A. Kazakov$^1$\footnote{$^{\circ}$}{{kazakov@physique.ens.fr}}}

\

\centerline{$^1${\it  Laboratoire de Physique Th\'eorique de l'Ecole
Normale Sup\'erieure,\footnote{$^\dagger$}{Unit\'e de Recherche du
Centre National de la Recherche Scientifique et de  l'Ecole Normale
Sup\'erieure et \`a l'Universit\'e de Paris-Sud.} }}
\centerline{{\ \ \ \it 24 rue Lhomond, 75231 Paris CEDEX, France}}
\centerline{$^2${\it Service de Physique Th\'eorique,
CNRS - URA 2306, C.E.A. - Saclay,}}
\centerline{  F-91191 Gif-Sur-Yvette CEDEX, France}
\centerline{$^3$ \it V.A.~Fock Department of
Theoretical Physics, St.~Petersburg
University, Russia}

\bigskip
%

%
%%%%%%%%%%%%%%%%%%%%%%%%%%%%%%%%%%%%%%%%%%%%%%%%%%%%%%%%%%%%%%%%%%%%%%%%%%
  \vskip 1cm
\baselineskip8pt{

\baselineskip12pt{
\noindent

We calculate the free energy, energy and entropy in the matrix quantum
mechanical formulation of 2D string theory in a background strongly
perturbed by tachyons with the imaginary Minkowskian momentum $\pm
i/R$ (``Sine-Liouville'' theory).  The system shows a thermodynamical
behaviour corresponding to the temperature $T={1/(2\pi R)}$.  We show
that the microscopically calculated energy of the system satisfies the
usual thermodynamical relations and leads to a non-zero entropy.  }}

%\draft
\Date{}

%\bigskip \bigskip \bigskip \bigskip

%if you want double-space, use e.g.
\baselineskip=14pt plus 2pt minus 2pt

%%%%%%%%%%%%%%%%%%%%%%%%%%%%%%%%%%%%%%%%%%%%
    \newsec{Introduction}
%%%%%%%%%%%%%%%%%%%%%%%%%%%%%%%%%%%%%%%%%%%%

Nontrivial backgrounds in the string theory often have a
thermodynamical behaviour corresponding to a certain temperature and a
classically big entropy.  Typical examples are the black holes where
the thermodynamics manifests itself by the Hawking radiation at a
certain temperature $T_H$ and the Bekenstein-Hawking entropy.  The
latter is expressed through the classical parameters of the system
(the area of the horizon) and satisfies the standard thermodynamical
relations with the free energy and energy (mass of the black
hole).\foot{For review of black hole physics see
\refs{\KumVas,\DasMathur}.}

In this paper we will try to demonstrate that the thermodynamical
behaviour with a temperature $T$ is a rather natural phenomenon in the
2-dimensional string theory in specific backgrounds created by tachyon
sources with Euclidean momenta corresponding to the Matsubara
frequencies $2\pi k/T, k=1,2,\ldots$.  For $k=1$ this system is T-dual
to the so called ``Sine-Liouville'' theory conjectured to describe 2D
string theory on the Euclidean black hole background \FZZ.  The latter
can be obtained introducing a vortex source with the Euclidean time
compactified on a radius $R=1/2\pi T$. It was studied in \KKK\ and
further in \refs{\AK,\IK} in the matrix quantum mechanical (MQM)
formulation using the integrability properties.

We will use here the MQM in the singlet sector to study this system in
Minkowskian time, perturbed by a source of ``tachyons'' with imaginary
momenta. This approach, elaborated in the paper \AKK, using some early
ideas of \DMP, extensively relies on the classical Toda integrability
and the representation of the system in terms of free fermions arising
from the eigenvalues of the matrix field.  We will demonstrate that
the microscopically defined energy of the system, calculated as the
energy of (perturbed) Fermi sea, coincides with its
thermodynamical counterpart $E$ calculated as the derivative of the
free energy with respect to the temperature.  The free energy itself
can be found from the Fermi sea due to its relation with the number of
particles $N={\p\CF/ \p \mu}$.

This coincidence of the microscopic and thermodynamical energies looks
satisfactory, though a little bit mysterious to us. We could not
figure out here where the entropy, also following from this
derivation, comes from since the system looks like a single classical
state of the moving Fermi liquid. The formulas become especially
simple in the black hole limit proposed in \KKK, where the energy and
entropy dominate with respect to the free energy.  On the other hand,
the entropy naturally disappears in the opposite limit of the trivial
linear dilaton background where the perturbation source is absent.

The entropy seems to be more visible in the T-dual formulation of the
theory based on the vortex perturbation in the compact Euclidean time
where we deal with the higher representations of the $SU(N)$ symmetry
of the MQM \KZ. However the direct counting of the nonsinglet states
appears to be technically a difficult problem. But we still think that
our observation could shed light on the origin of the entropy of the
black hole type solutions in the string theory.

The paper is organized as follows.  In the next section we review the
fermionic representation of $c=1$ string theory. In section 3 we show
how one can describe the tachyon perturbations in this
formalism. Section 4 presents the explicit solution for the particular
case of the Sine-Liouville theory.  In section 5 we evaluate the grand
canonical free energy and energy of the system using the techniques of
integration over the Fermi sea.  In section 6 we demonstrate that the
quantities found in the previous section satisfy the thermodynamical
relations and give rise to a nonvanishing entropy. The parameter $R$
of the perturbations can be interpreted as the inverse
temperature. Finally, we discuss our results and remaining problems.

%%%%%%%%%%%%%%%%%%%%%%%%%%%%%%%%%%%%%%%%%%%%%%%
\newsec{The  $c=1$ string theory as the collective theory of the Fermi sea}
%%%%%%%%%%%%%%%%%%%%%%%%%%%%%%%%%%%%%%%%%%%%%%%

The $c=1$ string theory can be formulated in terms of collective
excitations of the Fermi sea of the upside-down harmonic oscillator 
\refs{\BRKA,\PARISI,\GRMI,\GIZI}.
The tree-level tachyon dynamics of the string theory is contained in the
semiclassical limit of the ensemble of fermions, in which the
fermionic density in the phase space is either one or zero.  Each
state of the string theory corresponds to a particular configuration
of the incompressible Fermi liquid \refs{\POLCHINSKI,\KLEBANOV,\JEVICKI}.

In the quasi-classical limit the motion of the fermionic liquid is
determined by the classical trajectories of its individual particles
in the phase space.  It is governed by the one-particle Hamiltonian
\eqn\freeHV{
H_0= \hf\left( p^2 -x^{2}\right), } where the coordinate of the
fermion $x$ stands for an eigenvalue of the matrix field and
$p=-i{\p\over\p x}$ is its conjugated momentum.  We introduce the
chiral variables
\eqn\lcvar{
\xpm (t)= {x (t) \pm p(t)  \over \sqrt{2}},}
and define the Poisson bracket as $\{f,g\}={\p f \over \p \xm
}{dg\over \p \xp} - {\p f \over \p \xp }{dg\over \p \xm} $.  In these
variables $H_0 = -\xp\xm$ and the equations of motion
%\eqn\freeeq{ \dot \xpm = \{ \xpm,H_0\} = \pm \xpm}
have a simple solution
\eqn\freesol{
\xpm (t) = e^{\pm t} \zpm,}
where the initial values $z_{\pm}$ parameterize the points of the Fermi
sea.  Each trajectory represents a hyperbole $H_{0}(p, x) = E.$ The
state of the system is completely characterized by the profile of the
Fermi sea that is a curve in the phase space which bounds the region
filled by fermions. For the ground state, the Fermi sea is made by the
classical trajectories with $E<-\mu$, and the profile is given by the
hyperbole
\eqn\hyperx{ \xp \xm= \mu.}
The state is stationary since the Fermi surface coincides with one of
the classical trajectories and thus the form of the Fermi sea is
preserved in time.  For an arbitrary state of the Fermi sea, the Fermi
surface can be defined more generally by
\eqn\eqpr{ \xp\xm = M(\xp, \xm).}
It is clear from \freesol\ that a generic function $M$ in \eqpr\ leads
to a time-dependent profile. However, this dependence is completely
defined by \freesol\ and it is of little interest to us. We can always
replace \eqpr\ by the equation for the initial values $\zpm$.

For a given profile, we define the energy of the system and the
number of fermions as the following  integrals over the Fermi sea
\def\EO{E_0}
\eqn\ener{
\EO= \dpi\mathop{\int\int}\limits_{\fs} dx\,dp\ H_0(x,p), \qquad
N= \dpi \mathop{\int\int}\limits_{\fs} dx\, dp .  
}
It is implied that the integrals are bound by a cut-off
at a distance $\sqrt{2\Lambda}$.
For example, we can restrict the integration to $0<z_{\pm}<\sqrt{\Lambda}$.
For the ground state \hyperx, dropping non-universal terms
proportional to the cut-off one reproduces the well known result \KAZMIG
\eqn\enerfr{
\EO(\mu)= -   {1\over 4\pi}
\mu^{2}\log (\mu/\Lambda), \qquad
N(\mu) =  {1\over 2\pi} \mu\log (\mu/\Lambda).
}

Given the number of fermions, one can also introduce the grand canonical
free energy through the relation
\eqn\dFdm{
N = \p \CF/\p \mu,}
where $\CF$ is given in terms of the partition function
for a time interval $T$ by
\eqn\FRZ{
\CF = -\log \CZ/T.}
For the case of the ground state, one finds from \enerfr\ that the
(universal part of the) grand canonical free energy is related to the
energy of the fermions as
\eqn\FEN{ \CF  = \EO  +\mu  N  = {1\over 4\pi} \mu^{2}\log(\mu/\Lambda).}

%%%%%%%%%%%%%%%%%%%%%%%%%%%%%%%%%%%%%%%%%%%%%%%%%
\newsec{The profile of the Fermi sea for time dependent
tachyon backgrounds}
%%%%%%%%%%%%%%%%%%%%%%%%%%%%%%%%%%%%%%%%%%%%%%%%

The ground state \hyperx\ describes the simplest, linear dilaton
 background of the bosonic string.  Now we would like to study more
 general backgrounds characterized by condensation of tachyons with
 nonzero momenta.  Such backgrounds correspond to time-dependent
 profiles \eqpr, characterized by the function $M(\zp,\zm)$.  We
 restrict ourselves to the states that contain only tachyon
 excitations whose momenta belong to a discrete lattice
\eqn\spectr{ p_n= i n/R,  \quad n\in \Z.}
They look as  Matsubara frequencies corresponding to the temperature
$T=1/(2\pi R)$, or to the compactification in the Euclidean time with
the radius $R$.  We will perturb the system in the same manner as in
\AKK. It has been shown that at the quasiclassical level
the perturbations can be completely characterized by the asymptotics
of the profile of the Fermi sea at $\zp \gg \zm$ and $\zm
\gg \zp$. Accordingly, the equation for the profile can be written in
two forms which should be compatible with each other \AKK
\eqn\zas{\zp\zm =\Mpm(\zpm)= \sum_{k=1}^n k t_{\pm k} \zpm^{k/R}
 + \mu + \sum\limits_{k=1}^{\infty} v_{\pm k} \zpm^{-k/R}, } where
$t_{\pm k}$ are coupling constants associated with the perturbing
operators and defining the asymptotics of the Fermi sea for $z_+\gg
z_-$ and $z_-\gg z_+$ correspondingly.  $v_{\pm k}$ are the
coefficients to be found, which completely fix the form of the
profile.

It was shown in \AKK\ that
such perturbations are described by the dispersionless limit of
a constrained Toda hierarchy. The phase space coordinates $\zpm$
play the role of the Lax operators, $t_{\pm k}$ are the Toda times,
and eq. \zas\ appears as the constraint (string equation) of
the hierarchy \IK.

The integrability allows to find the explicit solution for the profile.
It is written in the parametric form as follows
\eqn\zom{
\zpm(\o, \mu)= e^{-{1\over 2R}\chi(\mu)}\o^{\pm 1}
\left(1+\sum_{k= 1}^n a_{\pm k}(\mu)\ \o^{\mp k/R}\right),
}
where $\chi$ is the so called string susceptibility related to the
Toda $\tau$-function as $\chi=\p^2_\mu \log \tau$. The latter is
the generating function for the coefficients $v_{\pm k}$ in \zas
\eqn\pot{
v_{\pm k}=\p_{t_{\pm k}} \log \tau,}
which can be identified with one-point vertex correlators \refs{\AK,\AKK}.
Moreover, the change of variables $\zpm \to \log \o,\mu$ turns out to be
canonical, {\it i.e.},
\eqn\ccrom{ \{ \mu,\log \o  \} = 1. }
To find the coefficients $a_{\pm k}$ it is enough to use
a simple procedure suggested in \IK.
One should substitute the expressions \zom\ in the profile
equations \zas\ and compare the coefficients in front of $\o^{\pm k/R}$.

%%%%%%%%%%%%%%%%%%%%%%%%%%%%%%%%%%%%%%%%%%
\newsec{ Solution for Sine-Gordon coupled to gravity}
%%%%%%%%%%%%%%%%%%%%%%%%%%%%%%%%%%%%%%%%%

Let us restrict ourselves to the case $n=1$ corresponding to the
Sine-Gordon theory coupled to gravity
or the so called Sine-Liouville theory.
In this case there are only $\tpm$ coupling constants and
the equation \zom\ takes the form \AK\
\eqn\zomSG{
\zpm = e^{-{1\over 2R}\chi} \o^{\pm 1} (1+a_\pm\o^{\mp \oR} ),
}
where $\chi$ can be found from \refs{\KKK,\MOORE}
\eqn\coefzom{
\mu e^{\oR \chi} + (\oR-1)\tp\tm e^{{2R-1 \over R^2}\chi} =1,
\qquad
a_\pm = \tmp e^{{2R-1 \over 2R^2}\chi}.  } Assuming that $\hf<R<1$
(which means that the corresponding compactification radius is between the
Kosterlitz-Thouless and the self-dual one), the first equation for the
sucseptability can also be rewritten in terms of scaling variables:
\eqn\param{
w=\mu\xi, \qquad
\xi=\left(\lambda\sqrt{1/R-1}\right)^{-{2R\over 2R-1}},
}
where $\lambda=\sqrt{\tp\tm}$.
The result reads
\eqn\eqF{\eqalign{
\chi&=R\log \xi+\X(w), \cr
w &=\ef-\eg.
}}

%%%%%%%%%%%%%%%%%%%%%%%%%%%%%%%%%%%%%%%%%%%%%%%%%%%%%%%%%%%%%%%%%%%%%
\newsec{Free energy and energy of the perturbed background}
%%%%%%%%%%%%%%%%%%%%%%%%%%%%%%%%%%%%%%%%%%%%%%%%%%%%%%%%%%%%%%%%%%%%%

Let us consider the average over the Fermi sea of an observable $\CO$
defined on the phase space
\eqn\integV{
<\CO>=\dpi\mathop{\int\int}\limits_{\fs} d\zp d\zm\,
\CO(\zp,\zm). }
Here the boundary of the integration is defined by the profile equation
\zas. Since the change of variables described
by eq. \zom\ is canonical (see eq. \ccrom), one can write
\eqn\avV{
<\CO>=\dpi\int\limits_{\mu} ds \int\limits_{\bom(s)}^{\bop(s)}
{d\o \over \o}\,  \CO(\zp(\o,s),\zm(\o,s)).
}
The limits of integration over $\o$ are defined by the cut-off.
We choose it as two walls at
$\zp=\sqrt{\Lambda}$ and $\zm=\sqrt{\Lambda}$.
Then the limits can be found from the equations
\eqn\grcon{
\zpm(\bopm(s),s)=\sqrt{\Lambda}.
}

First, consider the case of $\CO=1$. Then the integral \integV\
gives the number of fermions in the Fermi sea,
i.e., $N=<1>$.
Taking the derivative with respect to $\mu$, we obtain
\eqn\numb{
\p_{\mu}N=-\dpi \log{\bop(\mu)\over \bom(\mu)}.}
In this case it is enough to restrict ourselves to the main
order  in the cut-off $\Lambda$  for the boundary values $\bopm$.
 From \grcon\ and \zom\ we find to this order
\eqn\grsolf{
\bopm=\oo^{\pm 1}, \qquad \oo=\sqrt{\Lambda e^{\chi/R}}.
}
Combining \numb\ and \grsolf, we find
\eqn\frener{
\p_{\mu}N=
-\dpi \log\Lambda- {1 \over 2\pi R} \chi.  } Taking into account the
relation \dFdm, we find that the free energy coincides with the
logarithm of the $\tau$-function and $\beta =2\pi R$ represents the
time interval at which the theory is compactified
\eqn\frenercom{
\CF=- {1\over\beta }\log \tau.  }

This derivation of the grand canonical free energy was done in \AKK.
However, using the general formula \avV, we can also find the energy
of the system, at least for the case of the Sine-Liouville theory.
Since it is given by the integral of the free Hamiltonian \ener, from
\avV\ and
\zomSG\ one obtains
\eqn\freeav{
\p_{\mu}E=-\p_{\mu}<\zp\zm>=\left. \dpi
e^{-\oR\chi}\left\{ (1+a_+a_-)\log \o -
R a_+\o^{-1/R}+R a_-\o^{1/R}\right\}\right|_{\bom}^{\bop}.
}
To find the final expression one needs to know
the limits $\bopm$ in the second order in the cut-off.
It is easy to show from \grcon\ that eq. \grsolf\ should be changed by
\eqn\grom{
\bopm=\oo^{\pm 1}(1\mp a_{\pm}\oo^{-1/R}).
}
Substituting these limits into \freeav, one finds
\eqn\freeres{
\p_{\mu}E =
{1\over 2 \pi } e^{-{1\over R}\chi}\left\{ (1+a_+a_-)
\left(\log\Lambda+\oR\chi\right) -2a_+a_-\right\}+
{R\over 2\pi}(\tp+\tm)\Lambda^{1/2R}.
}
Integrating over $\mu$ and using
$a_+a_-={R\over 1-R} e^{{2R-1\over R^2}\X}$,  we get
\eqn\valen{\eqalign{
2\pi E & =
\xi^{-2}\left(
{1\over 2R} \ea +{2R-1 \over R(1-R)}\eb- {1\over 2(1-R)}\ec \right)
(\chi+R\log\Lambda) \cr & + \xi^{-2} \left({1\over 4} \ea-{R \over
1-R}\eb+ {R(4-5R)\over 4(1-R)^2}\ec\right)+R(\tp+\tm)\mu
\Lambda^{1/2R}.  }} 
We observe that the last term is non-universlal since it does not contain
a singularity at $\mu=0$. However, $\log \Lambda$ enters in a
non-trivial way. 
It is combined with a non-trivial function of $\mu$ and $\xi$.

%%%%%%%%%%%%%%%%%%%%%%%%%%%%%%%%%%%%%%%%%%%%%%%%%%%%%%%%%%%%%%%%%%%%%
\newsec{Thermodynamics of the system}
%%%%%%%%%%%%%%%%%%%%%%%%%%%%%%%%%%%%%%%%%%%%%%%%%%%%%%%%%%%%%%%%%%%%%

We found above the energy of the Sine-Liouville theory \valen\
and also restored the free energy in the grand canonical ensemble
for a general perturbation.
In the Sine-Liouville case one can integrate the equation \eqF\
to get the explicit expression \KKK
\eqn\valfe{
2\pi \CF=-{1\over 2R} \mu^2 (\chi+R\log\Lambda)- \xi^{-2}\left(
{3\over 4} \ea-{R^2-R+1 \over 1-R}\eb+{3R\over 4(1-R)}\ec\right).
}

Now we are going to establish thermodynamical properties of our system.
It is expected to possess them since the spectrum of perturbations
corresponds to the Matsubara frequencies typical for a system at a
nonzero temperature.  First of all, note that in the standard
thermodynamical relations the free energy appears in the canonical,
rather than in grand canonical ensemble. Therefore we define
\eqn\legtr{
F=\CF-\mu{\p \CF \over \p \mu}.
}
It is easy to check that it can be written as
\eqn\trfr{\eqalign{
2\pi F&={1\over R}\int^{\mu} s\chi(s)ds=
{1\over 2R} \mu^2 (\chi+R\log\Lambda) \cr
& + \xi^{-2}\left(
{1\over 4} \ea-R\eb+{R\over 4(1-R)}\ec\right).
}}
Following the standard thermodynamical relations, 
let us also introduce the entropy as a difference of the energy and
the free energy:
\eqn\diff{
S=\beta (E-F).}
One finds from \valen\ and \trfr
\eqn\valent{\eqalign{
S&=\xi^{-2} \left(
{R\over 1-R}\eb-{1\over 2(1-R)}\ec \right)
(\chi+R\log\Lambda)
 \cr
& + \xi^{-2}\left( -{R^3\over 1-R}\eb +
{R^2(3-4R)\over 4(1-R)^2}\ec\right) +
{R^2\over 2\pi}(\tp+\tm)\mu\Lambda^{1/2R}.
}}

Thus, all quantities are supplied by a thermodynamical interpretation.
The role of the temperature is played by $T=\beta^{-1}={1/ 2\pi R}$.
However, to have a consistent picture, the following (equivalent)
thermodynamical relations should be satisfied:
\eqn\termrel{\eqalign{
S&=-{\p F \over \p T}=2\pi R^2{\p F \over \p R}, \cr
E&={\p (\beta F) \over \p \beta}={\p (R F) \over \p R}. \cr
}}
In these relations it is important which parameters are hold
to be fixed when we evaluate the derivatives with respect to $R$.
Since we use the free energy in the canonical ensemble,
one should fix the number of fermions $N$ and the coupling constant
$\lambda$ but not the chemical potential $\mu$.

Another subtle point is that {\it a priori}
the relation of our initial parameters to the corresponding
parameters which are to be fixed may contain
a dependence on the inverse temperature $R$. Therefore,
in general we should redefine
\eqn\redcon{\eqalign{
\xi & \longrightarrow  \xi=\left( a(R)\lambda \right)^{-{2\over 2-R}}, \cr
\Lambda & \longrightarrow  b(R)\Lambda, \cr
F & \longrightarrow F+\hf c(R)\mu^2.
}}
We find the functions $a(R)$, $b(R)$ and $c(R)$ from the condition
\termrel, where $N$, $\lambda$ and $\Lambda$ are fixed.
We emphasize that it looks rather non-trivial to us that it is
possible to find such functions of $R$ that this condition is
satisfied.  It can be considered as a remarkable coincidence
indicating that the functions $F$, $E$ and $S$ do have the
interpretation of thermodynamical quantities as free energy,
energy and entropy correspondingly.  In appendix A it is shown that
the parameters should be taken as
\eqn\PAR{
a(R)=\left({1-R\over R^3}\right)^{1/2}, \qquad
b(R)=1, \qquad c(R)=0.}

It is worth noting several features of this solution. First of all,
the plausible property is that we do not redefine the cut-off and the
free energy. On the other hand, the constant $\lambda$ coincides with
its initial definition \param\ if to absorb one $R$ factor into
definition of the coupling constants \foot{This corresponds to the
fact that one should not introduce such $R$ factor in front of the
perturbing potential as was done in \KKK.} 
$\tpm$ what probably indicates the right
relation of $\lambda$ with the corresponding coupling constant in
the conformal Sine-Liouville theory.  Another important consequence of
the choice \PAR\ is that the entropy \valent\ is proportional to
$\lambda^2$ and therefore vanishes in the absence of perturbations.
It appears to be non-zero only when the background of the string
theory is described by a time dependent Fermi sea.  However, a
puzzle still remains how to calculate the entropy in a similar way as
the energy was evaluated, that is by directly identifying the
corresponding degrees of freedom which give rise to the same
macroscopic state and, hence, the entropy.

%%%%%%%%%%%%%%%%%%%%%%%%%%%%%%%%%%%%%%%%%%%%%%%%%%%%%%%%%%%%%%%%%%%%%
\subsec{The "Black hole" limit}
%%%%%%%%%%%%%%%%%%%%%%%%%%%%%%%%%%%%%%%%%%%%%%%%%%%%%%%%%%%%%%%%%%%%%

Let us restrict ourselves to the limit $\mu\rightarrow 0$.  If we
perturbed the 2D string theory by the vortex, instead of the tachyon,
modes it would correspond to the black hole limit \refs{\FZZ,\KKK}.
In this limit we have $X=0$.  As a result, we obtain from \trfr,
\valen\ and \valent\ for
all three thermodynamical quantities:
\eqn\valallbh{\eqalign{
2\pi F&={(2R-1)^2\over 4(1-R)}\tl^{4R\over 2R-1} , \cr
2\pi E&={2R-1\over 2R(1-R)}\left( \log(\Lambda\xi)
-{R\over 2(1-R)}\right) \tl^{4R\over 2R-1} , \cr
S&={2R-1\over 2(1-R)}\left(\log(\Lambda\xi)-{R^2(3-2R)\over 2(1-R)}\right)
\tl^{4R\over 2R-1} , \cr
}} where $\tl=a(R)\lambda$.  It is worth to note an interesting
feature of this solution.  Since $\log(\Lambda\xi)\gg 1$, we get $2\pi
E \approx S$ and in comparison with these quantities $F$ is negligible.  
This is to be compared with the result of \NP\ that the
analysis of the dilatonic gravity derived from 2D string theory leads
to the vanishing free energy and to the equal energy and entropy.  We
see that it can be true only in a limit, but the free energy can never
be exactly zero since it is its derivative with respect to the
temperature that produces all other thermodynamical quantities.

For the special case of $R=2/3$ which is T-dual to
the stringy black hole model, we find:
\eqn\allbh{\eqalign{
\beta F&={1\over 18}\tl^{8} , \cr
\beta E&=\left({1\over 2}\log(\Lambda\xi)-{1\over 2}\right)\tl^{8}, \cr
S&=\left({1\over 2}\log(\Lambda\xi)-{5\over 9}\right)\tl^{8}. \cr
}}
For the dual compactification radius $R=3/2$, all quantities change their
sign:
\eqn\allbh{\eqalign{
\beta F&=-{3}\tl^{3} , \cr
\beta E&=-\left( {2}\log(\Lambda\xi)-{3}\right)\tl^{3}, \cr
S&=-{2}\log(\Lambda\xi)\tl^{3}. \cr
}}
This might be interpreted as a result for the background dual to
the black hole of a negative mass, which does appear in the analysis of \DVV.

%%%%%%%%%%%%%%%%%%%%%%%%%%%%%%%%%%
\newsec{ Conclusions and problems}
%%%%%%%%%%%%%%%%%%%%%%%%%%%%%%%%%

%The main puzzle remained in our approach sounds in a familiar way to
%the black hole physicists: where are the states corresponding to the
%entropy of a classical gravitating object. This object is usually a
%black hole, and in our case it is a nontrivial tachyonic background
%conjectured to be T-dual to the dilatonic black hole? 

The main puzzle of the black hole physics is 
to find the states corresponding to the
entropy of a classical gravitating system. In our case, instead 
of a black hole, the role of such system is played by
a nontrivial tachyonic background. However, it manifests the similar 
properties as the usual dilatonic black hole, which appears as a 
T-dual perturbation of the initial flat spacetime.
Therefore, it is not surprising that the studied background
also possesses a nonvanishing entropy.

In our case we have a little more hope to track the origin of the entropy,
than, for example, in the study of the Schwarzschild black hole, 
since we are dealing with the well defined string theory and can identify its
basic degrees of freedom. It allowed us in this paper to find
microscopically the energy and free energy of the
``Sine-Liouville'' background. The entropy was calculated not
independently, but followed from the basic thermodynamical relations.
Our main result is the coincidence of this microscopic and the
thermodynamical energies. The latter is calculated as a derivative of the free
energy with respect to the temperature. Note that the ``temperature''
appears from the very beginning as a parameter of the tachyonic
perturbation, and not as a macroscopic characteristic of a heated
system. That is why the thermodynamical interpretation comes as a
surprise. 

An analogous situation can be seen in the Unruh effect \Unruh. There the
temperature seen by the accelerating observer is defined by the value
of the acceleration, which does not {\it a priory} have any
relation to thermodynamics.
In fact, in the latter case the temperature can be found from the  
analysis of the quantum field theory on the Rindler space
which is the spacetime seen by the accelerating observer.
The same should be true for our case. It should be possible to reproduce
the global structure of the tachyonic background 
from the matrix model solution, and then the thermodynamical 
interpretation should appear from the analysis of 
the spectrum of particles detected by some natural observer \SAprog.

One might try to find a hint to the thermodynamical behaviour of our
system in the very similar results for the T-dual system of \KKK\
where instead of tachyons we have the perturbation by vortices. There
the Euclidean time is compactified from the very beginning and the
Gibbs ensemble is thus explicitly introduced with $R$ as a
temperature parameter.  On the other hand, the coincidence of
microscopic energies in two mutually dual formulations (up to duality
changes in parameters, like $R\to 1/R$) is far from obvious. Even how
to find the energy in this dual case is not clear.

In conclusion, we hope that our observation will help to find a
microscopic approach to the calculation of entropy at least in the 2D
string theory with nontrivial backgrounds. Together with the plausible
conjecture that the dual system describes the 2D dilatonic black hole, 
it could open a way for solving the puzzle of the microscopic origin of
entropy in the black hole physics.

\bigskip
\noindent{\bf Acknowledgements:}
The authors are grateful I. Kostov for valuable discussions.
Also we thank the Theoretical Physics Division of CERN, 
where a part of the work was done,
for the kind hospitality. This work of S.A. and V.K. was partially
supported by European Union under the RTN contracts HPRN-CT-2000-00122
and -00131. The work of S.A. was supported  in part by
European network EUROGRID HPRN-CT-1999-00161.

%%%%%%%%%%%%%%%%%%%%% Appendicies %%%%%%%%%%%%%%%%%%%%%%%%%%%%%%%%

%%%%%%%%%%%%%%%%%%%%%%%%%%%
\appendix{A}{Calculation of $\p F/\p R$}
%%%%%%%%%%%%%%%%%%%%%%%%%%%

In this appendix we calculate the derivative of
the free energy \trfr\ with respect to $R$.
First, note that due to \legtr\ and \dFdm, we have
\eqn\derF{
\left({\p F \over \p R}\right)_{N,\lambda,\Lambda}=
\left({\p \CF \over \p R}\right)_{\mu,\lambda,\Lambda}=
\left({\p \CF \over \p R}\right)_{\mu,\xi,\Lambda}+
\left({\p \xi \over \p R}\right)_{\lambda}
\left({\p \CF \over \p \xi}\right)_{\mu,\Lambda,R}.
}
From \valfe\ and \redcon\ one can obtain
\eqn\derFxi{
2\pi \xi \left({\p \CF \over \p \xi}\right)_{\mu,\Lambda,R}=
-\xi^{-2}{R(2R-1)\over 1-R}\left( \eb-{1\over 2R}\ec\right),
}
\eqn\derxi{
\xi^{-1} \left({\p \xi \over \p R}\right)_{\lambda}=
-{1\over R(2R-1)}\left(\log\xi +2R^2{d \log a(R) \over d R}\right).
}
The most complicated contribution comes from the first term in \derF.
From \valfe, \redcon\ and \eqF\ one finds
\eqn\derFR{\eqalign{
2\pi \left({\p \CF \over \p R}\right)_{\mu,\xi,\Lambda} &=
\xi^{-2}\left({1\over 1-R}\eb -{1\over 2R(1-R)}\ec\right){\X \over R}\cr
&+\xi^{-2}\left( {R(2-R)\over (1-R)^2}\eb -{3\over 4(1-R)^2}\ec \right) \cr
&-\hf \mu^2 \left({d\log b(R)\over dR}+{d c(R)\over dR}\right) .
}}
Putting the results \derFxi, \derxi\ and \derFR\ together, we obtain
\eqn\derFFR{\eqalign{
2\pi\left({\p F \over \p R}\right)_{N,\lambda,\Lambda}&=
  {1\over R^2} \xi^{-2}\left({R\over 1-R}\eb -
{1\over 2(1-R)}\ec\right) \chi  \cr
&+\xi^{-2}\left( {R(2-R)\over (1-R)^2}\eb
-{3\over 4(1-R)^2}\ec \right) \cr
&+\xi^{-2}\left({2R^2\over 1-R}\eb -{R\over 1-R} \ec
\right){d \log a(R) \over d R} \cr
&-\hf \mu^2 \left({d\log b(R)\over dR}+{d c(R)\over dR}\right) .
}}
We observe that the term proportional to the susceptibility $\chi$
coincides with the corresponding term in the entropy \valent\
what can be considered as the main miracle of the present derivation.
To get the coincidence of other terms, we obtain a system of three
equations, from which only two are independent:
\eqn\EQforPar{\eqalign{
  {d\log b(R)\over dR} +{d c(R)\over dR} & =0,  \cr
  2{d \log a(R) \over d R}-{1\over R^2}\log b(R) & ={2R-3\over R(1-R)}. \cr
}}
Since we have only two equations on three functions, there is an ambiguity
in the solution. It can be fixed if we require that the entropy vanishes
without perturbations. Then we come to the following result
\eqn\PARA{
a(R)=\left({1-R\over R^3}\right)^{1/2}, \qquad
\log b(R)=0, \qquad c(R)=0,
} which ensures the thermodynamical relations \termrel.  Note that in
fact they are fulfilled only up to terms depending on the cut-off
$\log \Lambda$.  Such terms can not be reproduced by differentiating
with respect to the temperature, but they are considered as
non-universal in our approach and thus can be dropped.

 \listrefs

\bye